\documentclass[runningheads,orivec]{llncs}
\usepackage{graphicx}
\usepackage{diagbox}
\usepackage{xcolor}
\usepackage{makeidx} 
\usepackage[colorlinks,linkcolor=blue]{hyperref}
\usepackage{amsmath}
\usepackage{booktabs}
\usepackage[misc]{ifsym} 
\usepackage{bbding}
\usepackage{amsfonts}
\usepackage{multirow}
\usepackage{xspace}
\usepackage{makecell}
\usepackage{enumitem} 
\newcommand{\ourmodel}{MTCNet}

\begin{document}

\title{MTCNet: Motion and Topology Consistency Guided Learning for Mitral Valve Segmentation in 4D Ultrasound}

\titlerunning{Mitral Valve Segmentation in 4D Ultrasound}

\author{Rusi Chen\inst{1}\thanks{Rusi Chen, Yuanting Yang, Jiezhi Yao and Hongning Song contributed equally.\\
Qing Zhou and Xin Yang are the corresponding authors of this work.} 
\and Yuanting Yang\inst{2\star} 
\and Jiezhi Yao\inst{1\star} 
\and Hongning Song\inst{2\star} 
\and Ji Zhang\inst{2} 
\and Yongsong Zhou\inst{1} 
\and Yuhao Huang\inst{1} 
\and Ronghao Yang\inst{1} 
\and Dan Jia\inst{2}
\and Yuhan Zhang\inst{1} 
\and Xing Tao\inst{1} 
\and Haoran Dou\inst{3} 
\and Qing Zhou\inst{2}\textsuperscript{(\Letter)}
\and Xin Yang\inst{1}\textsuperscript{(\Letter)}
\and Dong Ni\inst{1,4,5,6}
} 

\institute{
\textsuperscript{$1$}Medical Ultrasound Image Computing (MUSIC) Lab, School of Biomedical Engineering, Medical School,
Shenzhen University, Shenzhen, China\\
\email{xinyang@szu.edu.cn} \\
\textsuperscript{$2$}Renmin Hospital of Wuhan University, Wuhan, China\\
\email{qingzhou.wh.edu@hotmail.com} \\
\textsuperscript{$3$}Centre for CIMIM, Manchester University, Manchester, UK \\
\textsuperscript{$4$}School of Artificial Intelligence, Shenzhen University, Shenzhen, China\\
\textsuperscript{$5$}National Engineering Laboratory for Big Data System Computing Technology, Shenzhen University, Shenzhen, China\\
\textsuperscript{$6$}School of Biomedical Engineering and Informatics, Nanjing Medical University, Nanjing, China\\
}

\authorrunning{R.Chen et al.}

\maketitle

\begin{abstract} 
Mitral regurgitation is one of the most prevalent cardiac disorders.
Four-dimensional (4D) ultrasound has emerged as the primary imaging modality for assessing dynamic valvular morphology. 
However, 4D mitral valve (MV) analysis remains challenging due to limited phase annotations, severe motion artifacts, and poor imaging quality.
Yet, the absence of inter-phase dependency in existing methods hinders 4D MV analysis.
To bridge this gap, we propose a Motion-Topology guided consistency network (\textbf{\ourmodel}) for accurate 4D MV ultrasound segmentation in semi-supervised learning (SSL). 
\ourmodel~requires only sparse end-diastolic and end-systolic annotations.
First, we design a cross-phase motion-guided consistency learning strategy, utilizing a bi-directional attention memory bank to propagate spatio-temporal features. This enables \ourmodel~to achieve excellent performance both per- and inter-phase.
Second, we devise a novel topology-guided correlation regularization that explores physical prior knowledge to maintain anatomical plausibility. Therefore, \ourmodel~can effectively leverage structural correspondence between labeled and unlabeled phases.
Extensive evaluations on the first largest 4D MV dataset, with 1408 phases from 160 patients, show that \ourmodel~performs superior cross-phase consistency compared to other advanced methods (Dice: 87.30\%, HD: 1.75mm).
Both the code and the dataset are available at \url{https://github.com/crs524/MTCNet}.

\end{abstract}

\section{Introduction}
Mitral regurgitation (MR) is a common cardiovascular disease with high morbidity and mortality~\cite{el2018mitral,mcdonagh20212021}.
Transesophageal Echocardiography (TEE) is the gold standard for diagnosing and quantifying MR. It offers a real-time view of the mitral valve (MV), providing both temporal and spatial perspectives~\cite{vahanian20222021}.
Accurate 4D MV segmentation enables precise measurement of MV structure and functional analysis, as well as patient-specific 3D printing for surgical planning.

\begin{figure}[!t]
	\begin{center}
    \includegraphics[width=1\columnwidth]{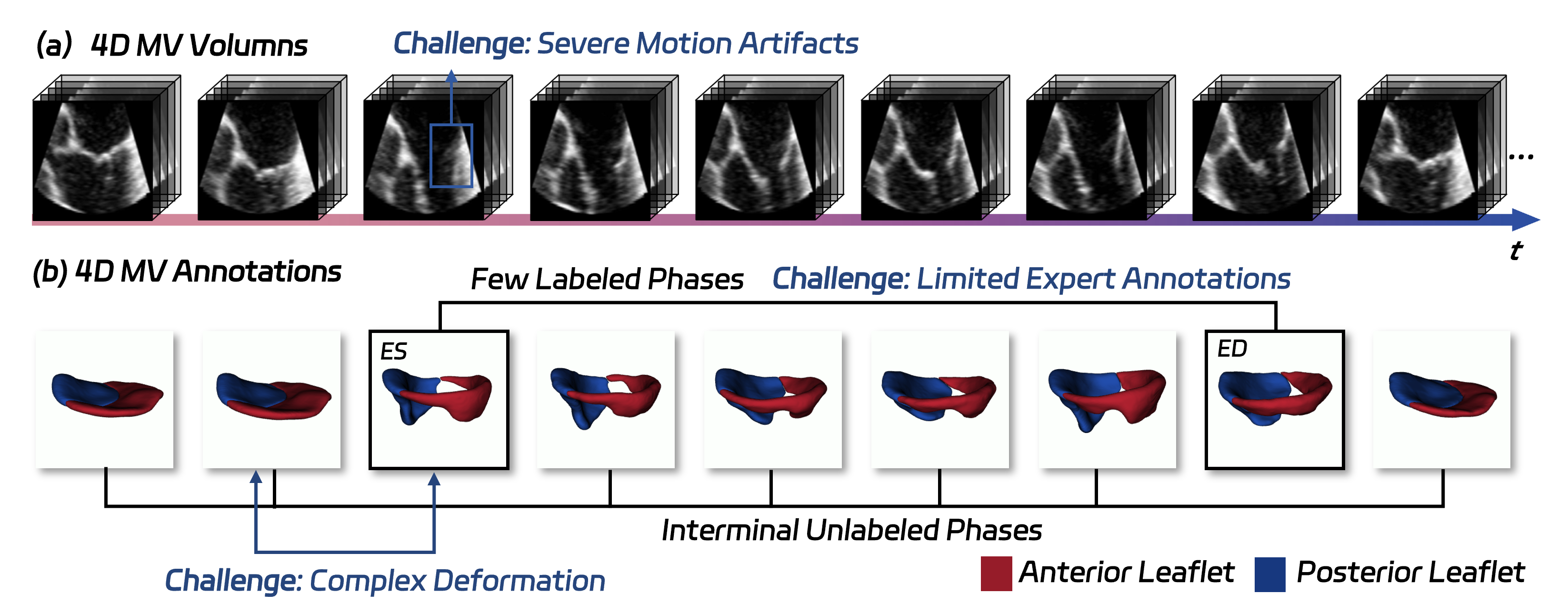}
	\end{center}
    \caption{Illustration of MV volumes and annotations in 4D ultrasound.}
	\label{fig:task}
\end{figure}

However, automatic MV segmentation in 4D ultrasound faces challenges caused by a few phase annotations, severe motion artifacts, and complex deformations (See Fig. \ref{fig:task}). 
To tackle these challenges, we aim to develop a 4D MV segmentation method with only end-diastolic (ED) and end-systolic (ES) phase annotations, while ensuring high accuracy and temporal coherence. 

Despite the promising progress in deep learning~\cite{yan2024foundation,luo2025tumor,huang2024segment,huang2025flip}, research on 4D MV segmentation remains limited, with most studies focusing on single-volume segmentation.
Some studies \cite{carnahan2021deepmitral,aly2022fully} employed U-Net \cite{ronneberger2015u} for 3D MV segmentation, while others \cite{chen2023automatic,munafo2024deep} utilized the nnU-Net \cite{isensee2021nnu} framework.
Although these methods achieve promising accuracy in static 3D volumes, they struggle with MV motion due to a lack of information extraction from unannotated phases.
Semi-supervised learning (SSL) \cite{EMAtarvainen2017mean,sslyu2019uncertainty,sslzhang2017deep} has emerged as a promising paradigm to enhance model performance with limited labeled data by utilizing the abundant information in unlabeled data.
Ivantsits et al. \cite{ivantsits2024mv} simplify MV as an ideal tubular sheet to reconstruct a 4D MV surface, which limits its use in simulating the common organic mitral regurgitation disease with valve abnormalities.
Munaf{`o} et al. \cite{munafo2025automatic} proposed an SSL method for 4D MV segmentation, where ED and ES phases of all patients are treated as labeled, while intermediate phases are considered unlabeled ones.
However, they treat each phase independently, neglecting patient-level phase consistency, leading to suboptimal segmentation.

To address these challenges, we leverage motion and topology-guided consistency learning to account for the relationships between patient-level labeled and unlabeled phases while preserving anatomical constraints.
In this paper, we propose a motion-topology guided consistency method termed \ourmodel~to regard all phases of a patient as input samples to mine cross-phase relevance.
Firstly, to structurally enhance the motion coherence, we introduce motion-guided consistency learning (MCL). Through a well-designed bi-directional memory bank (BMB), MCL can encourage \ourmodel~to effectively learn from key labeled phases and then propagate the semantic features to the unlabeled ones. 
Secondly, considering the anatomical prior of the MV's surface-volume invariance, we embed this property into a topology consistency regularization (TCR), optimizing inter-phase dissimilarity through an extra surface and volume continuity constraint.  Extensive evaluations on the largest 4D MV dataset demonstrate that \ourmodel~outperforms existing SSL approaches.
Our contributions are three-fold:
\begin{itemize}[label=\textbullet, itemsep=0.3em] 
    \item We introduce a novel motion and topology-guided learning framework for 4D MV segmentation, which achieves superior per-phase and inter-phase performance with sparse ES and ED annotations.
    \item We propose a MCL strategy to effectively propagate semantic information across labeled and unlabeled phases, thereby enhancing motion consistency.
    \item We design a simple yet effective TCR with surface and volume variance that boosts segmentation accuracy and preserves topological coherence.
\end{itemize}

\section{Methods}
\textbf{Problem Setting.}  
Given a patient-level sequence $\mathbf{D}$ comprising a labeled subset \( \mathbf{D}_l^i = \{ (X_l, Y, t) \} \) 
and an unlabeled subset
\( \mathbf{D}_u^i = \{ (X_u, t') \} \) for patient $i$, where $t$ and $t'$ represent the indices of the ES and ED phases, and the intermediate phases, respectively.
Our goal is to segment all phases within a cycle using a limited number of labeled phases. 
Specifically,  \( X_l \in \mathbb{R}^{D \times H \times W} \) and \( X_u \in \mathbb{R}^{D \times H \times W} \) denote labeled phases and unlabeled phases, respectively. \( Y \in \{0, 1\}^{D \times H \times W} \) represents the MV labels of $X_l$.

\subsection{Patient-level Cross-phase Learning Framework} 
\begin{figure}[!t]
	\begin{center}
    \includegraphics[width=1\columnwidth]{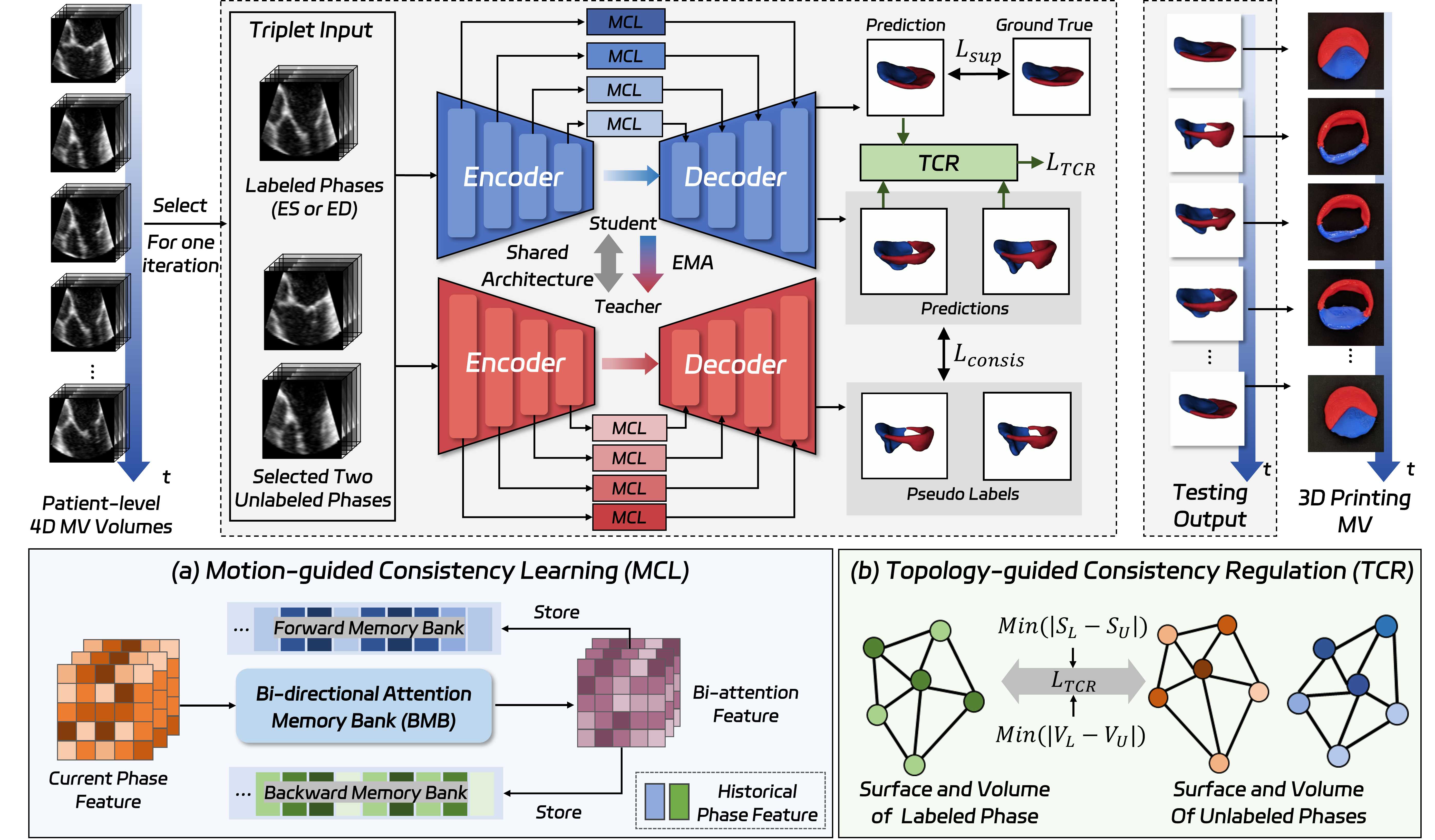}
	\end{center}
    \caption{Overall framework of our proposed \ourmodel.}
	\label{fig:framewok}
\end{figure}

The overall framework of our method is illustrated in Fig. \ref{fig:framewok}.
Inspired by the Mean Teacher architecture \cite{tarvainen2017mean}, a powerful SSL method, \ourmodel~is designed to enhance learning from both labeled and unlabeled data.
For each training iteration, \ourmodel~takes triplet phases $\mathcal{T}$ as input from the sequence $\mathbf{D}_l^i$. 
Specifically, one of the triplet phases is a labeled phase, while the other two phases are unlabeled volumes from intermediate phases. Thus, the input can be defined as $\mathcal{T} = \{ X_{l}, X_{u}^{1}, X_{u}^{2} \}$, where \(X_{l} \in \{X_{ES}, X_{ED}\}\).
Simultaneously, $\mathcal{T}$ is fed into the student and teacher models, which share the same architecture. 

During training, the teacher network supervises the student network by generating high-confidence pseudo-labels.
The teacher network's parameters $\theta_t$ are updated via exponential moving average (EMA) \cite{EMAtarvainen2017mean} from the student's parameters $\theta_s$.
It can be formulated as 
$\theta_t = \alpha \theta_t + (1 - \alpha) \theta_s$,
where $\alpha \in (0, 1)$ is the momentum mitigating the overfitting of the teacher network on limited labeled data.
Ultimately, \ourmodel~generates the predicted segmentation masks for both the labeled and unlabeled volumes. 
The total training objective is:
\begin{equation}
\label{segloss}
\mathcal{L}_{seg} = \mathcal{L}_{sup}(f_s(\theta), Y) + \beta \cdot \mathcal{L}_{consis}(f_s(\theta), f_t(\theta)),
\end{equation}
where $\beta$ is the loss weight, ${L}_{sup}$ and  ${L}_{consis}$ indicate the supervised loss and consistency loss, respectively. $f_s(\theta)$ represents the segmentation model's output of labeled volume, and  $f_t(\theta)$ for unlabeled volumes. Both losses combine dice and binary cross-entropy loss with a 0.8 to 0.2 weight ratio.
During the testing stage, \ourmodel~is able to predict masks for all phases in an end-to-end manner.

\subsection{Motion-guided Consistency Learning}
To effectively enhance the inter-phase motion coherence, we propose an MCL strategy through the BMB block. 
As shown in Fig. \ref{fig:framewok} (a), we propose forward memory bank $M_f$ and backward memory bank $M_b$ to store the temporal information for all phases of a patient. 
$M_f$ captures systolic-diastolic deformation, while $M_b$ encodes reverse motion, jointly modeling bi-directional patterns to reduce phase misalignment in cardiac cycles.

\begin{figure}[!t]
	\begin{center}
    \includegraphics[width=1\columnwidth]{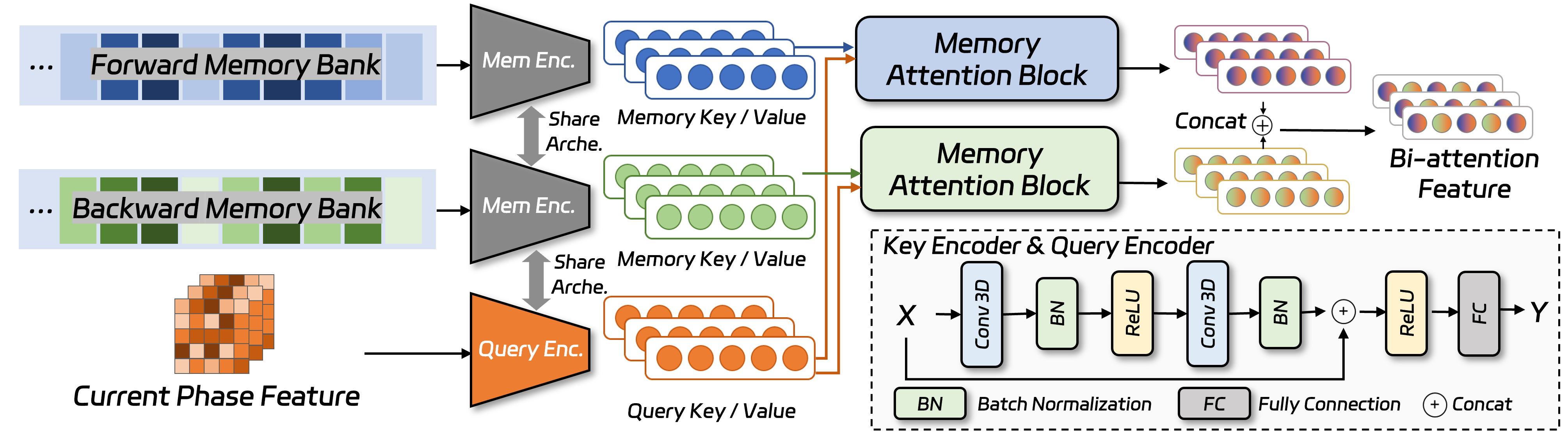}
	\end{center}
    \caption{Detailed design for bi-directional memory bank.}
	\label{fig:BMB}
\end{figure}

Fig. \ref{fig:BMB} illustrates the detailed design of the BMB block. 
Here, we take a memory bank to illustrate the mechanism, as both follow the same process. Given a memory bank $M$ containing $T$ memory phases, the memory encoder would generate memory key $k^M \in \mathbb{R}^{C^k \times T D H W}$ and memory value $v^M \in \mathbb{R}^{C^v \times T D H W}$. 
Similarly, the query encoder produces a query key $k^Q \in \mathbb{R}^{C^k \times D H W}$ and a query value $v^Q \in \mathbb{R}^{C^v \times D H W}$, where \( D \), \( H \), and \( W \) are the multi-scale feature dimensions. The current phase feature is the output feature of encoders, both in the student and teacher models.
Therefore, multi-scale memory aggregation minimizes information loss, especially for subtle boundary changes in low signal-to-noise ratio ultrasound phases.
Then, we compute the normalized affinity matrix $\mathbf{W}$ to effectively weigh the dependencies across phases:
\begin{equation}
    \mathbf{W}_{ij} = \frac{\exp(c(\mathbf{k}_i^M, \mathbf{k}_j^Q))}{\sum_n \exp(c(\mathbf{k}_n^M, \mathbf{k}_j^Q))},
\end{equation}
where $ \mathbf{W} \in \mathbb{R}^{T D H W \times D H W}$, $\mathbf{k}_i$ denotes the feature vector at the $i$-th position and $c$ represents the dot product. 

With the normalized affinity matrix $\mathbf{W}$, the aggregated readout feature $v^Q \in \mathbb{R}^{C^v \times D H W}$ for the query phase is computed as a weighted sum of the memory features using a top-k operation. 
To be specific, the weighted sum of the top-k memory features is calculated as:
$v_r^Q = v_r^M \mathbf{W}_r$,
where $r$ represents two forward and backward memory banks. 
Finally, the \( v_f^Q \) and \( v_b^Q \) are concatenated to obtain the bi-directional attention feature.
This feature is stored in the memory banks $M_f$ and $M_b$ and passed to the decoder to produce the segmentation mask.

\subsection{Topology-guided Consistency Regulation}
\label{sec:TCR}
To ensure structural integrity during complex deformations across phases, we propose the TCR mechanism. 
This approach is grounded in prior knowledge that surface area and volume should be stable throughout such deformations \cite{may1998constitutive} (see Fig. \ref{fig:framewok} (b)). 
Specifically, for a predicted probability map $P_t$ for $t$-th phase, we first obtain its binarized form $B_t = \mathbb{I}(P_t > 0.5)$, which represents the region of interest. 
Then, the normalized surface area $S(P_t)$ of the $t$-th phase can be defined as:

\begin{equation} 
S(P_t) =  \frac{\int_{\mathcal{S}} \nabla B_t(\mathbf{x}) , d\mathcal{A}(\mathbf{x})}{\int_{\mathcal{S}} \mathbb{I}(\nabla B_t(\mathbf{x}) > 0) , d\mathcal{A}(\mathbf{x}) + \epsilon} \approx  \frac{\sum_{v \in V} \nabla B_t(v) , \Delta A(v)}{\sum_{v \in V} \mathbb{I}(\nabla B_t(v) > 0) , \Delta A(v) + \epsilon},
\end{equation}
where $\nabla = (\partial_x, \partial_y, \partial_z)$ denotes 3D Sobel operators so that $\nabla B_t(\mathbf{x})$ represents the spatial gradient of the volumetric mask $B_t$ at location $\mathbf{x}$.
$\Delta A(v)$ is the discrete voxel-level form and $\epsilon$ prevents division by zero.
Sobel operators ensure the computation is differentiable, enabling seamless integration into gradient-based optimization frameworks.
Thus, the surface-area-based loss can be defined as:

\begin{equation}
\mathcal{L}_{\text{surf}} = \sum_{t=2}^3 \left( \left| 1 - \frac{S(P_t)}{S(P_1)} \right| + \lambda \left| S(P_t) - S(P_1) \right| \right).
\end{equation}

The dual-term design overcomes key limitations of static shape constraints. The relative term adapts to patient-specific anatomy via the annotated $S(P_1)$, while the absolute term prevents error propagation through physical consistency. 
Following a similar strategy to surface area consistency, the volume consistency loss can be formulated based on the voxel-wise summation of $B_t$:
\begin{equation} 
\mathcal{L}_{\text{vol}} = \sum_{t=2}^3 \left( \left| 1 - \frac{V(P_t)}{V(P_1)} \right| + \lambda \left| V(P_t) - V(P_1) \right| \right),
\end{equation}
where \( \sum_{i} B_t(i) \) is the voxel-wise summation of \( B_t \).
Finally, the total topological consistency regularization loss $\mathcal{L}_{\text{tcp}}$ can be written as:
$\mathcal{L}_{\text{tcp}} = \mathcal{L}_{\text{surf}} + \mathcal{L}_{\text{vol}}$.
The $\mathcal{L}_{\text{tcp}}$ acts as a physics-informed regularizer that enforces temporal plausibility.
Combined with equation \ref{segloss},
the total loss for triplet input phases integrates both segmentation and physiological regularization
$\mathcal{L}_{total} = \mathcal{L}_{\text{seg}} + \sigma \cdot \mathcal{L}_{\text{tcp}}$, where $\sigma$ is a constant set to 0.1 by default based on empirical observations.

\section{Experiments and Results}
\textbf{Implementation Details.}
For a fair comparison, all SSL experiments were conducted in the same setting, with only ES and ED annotated labels for training.
Our model was implemented with PyTorch 1.11.0 on two NVIDIA GeForce RTX A40 GPUs. 
All input volumes were all resampled to \(128 \times 128 \times 128\) based on isotropic spacing resizing. 
During training, the Adam optimizer was used with an initial learning rate of \(10^{-4}\), then reduced by 0.1 every 20 epochs.

\textbf{Datasets and Evaluation Metrics.}  
We collected 4D TEE MV data from 160 patients with 1408 phases from cooperating hospitals.
The in-house dataset comprised 147 cases with functional mitral regurgitation, and 13 normal cases.
The dataset was randomly split into training (112 cases), validation (16 cases), and testing (32 cases) subsets.
The number of phases varied per case due to heart rate differences, with an average of nine phases per cycle.
All imaging data were acquired using the Philips X5-1 transthoracic volume probe.
Two sonographers with more than 10 years of experience
manually annotated the MV volumes, consisting of anterior leaflets (AL) and posterior leaflets (PL).
We also validated the mid-diastolic (MD) phase and its adjacent transitional phase (MD-1) between MD and ED.
For evaluation, Dice coefficient, 95th-percentile Hausdorff Distance (HD), and Conformity (Conf) \cite{CHANG2009122}, were adopted for quantitative comparison.

\textbf{Comparison Study.}
We evaluated MTCNet against multiple SSL methods \cite{EMAtarvainen2017mean,sslyu2019uncertainty,Li_2020,luo2021semi,VERMA202290,10204429,huang2023complementary}, the registration-based approach VoxelMorph (VM) \cite{balakrishnan2019tmi}, and the two-shot video object segmentation method (T-VOS) \cite{yan2023two}. VM and T-VOS both require the first phase as reference. 
As depicted in Table~\ref{tab:results}, \ourmodel~achieves state-of-the-art (SOTA) performance, surpassing compared methods in Dice (\(87.30\%\)) and Conf (\(66.71\%\))  in all phases of segmentation. 
Specifically, our method surpasses the best baseline ICT (86.23\%) by 1.07\% (p$<$0.05).
Notably, although MD is the most challenging phase due to the leaflets are maximally separated and often near the left ventricular wall, \ourmodel~surpasses the sub-optimal T-VOS method by 2.4\% (\(82.69\%\) vs. \(80.29\%\)) in Dice, highlighting \ourmodel~'s ability in handling complex deformation. 
Additionally, compared to reference-phase-driven methods VM and T-VOS, \ourmodel~shows more potential results in capturing continuous motion more effectively.
Moreover, \ourmodel~achieves robust performance for both labeled phases (ES and ED) and unlabeled phases (MD and MD-1). 
This highlights \ourmodel~'s superior ability in per-phase and inter-phase performance.

\begin{table}[t]
\centering
\caption{Comparison results for different methods. All Phases stands for total phases in a cardiac cycle. * denotes statistically significant differences (t-test) between \ourmodel~and compared SOTA methods.}
\label{tab:results}
\scalebox{0.78}{
\begin{tabular}{l|c|c|c|c|c|c|c|c|c|c|c|c|c|c|c}
\Xhline{1.2pt}
\multirow{2}{*}{Method} & \multicolumn{3}{c|}{ES} & \multicolumn{3}{c|}{MD} & \multicolumn{3}{c|}{MD-1} & \multicolumn{3}{c|}{ED} & \multicolumn{3}{c}{All Phases} \\
\cline{2-16} 
                        & \makecell{Dice} & \makecell{HD} & \makecell{Conf} & \makecell{Dice} & \makecell{HD} & \makecell{Conf} & \makecell{Dice} & \makecell{HD} & \makecell{Conf} & \makecell{Dice} & \makecell{HD} & \makecell{Conf} & \makecell{Dice} & \makecell{HD} & \makecell{Conf} \\
\Xhline{1.2pt}
MT\cite{EMAtarvainen2017mean}               & 86.13 & 1.74  & 66.96 & 79.25 & 2.94  & 32.62 & 85.63 & 1.89  & 64.42 & 87.50 & 1.41  & 70.58 & 85.61\(^{*}\) & 1.78  & 63.27\(^{*}\) \\
UA-MT\cite{sslyu2019uncertainty}            & 86.44 & \textbf{1.68}  & 67.85 & 79.76 & 2.75  & 37.50 & 86.22 & 1.85  & 66.53 & 87.85 & 1.35  & 71.62 & 86.08\(^{*}\) & \textbf{1.69}  & 65.11\(^{*}\) \\
SASS\cite{Li_2020}                          & 85.90 & 1.85  & 66.45 & 79.67 & 2.74  & 35.21 & 86.64 & \textbf{1.69}  & 68.03 & 87.76 & 1.45  & 71.29 & 86.07\(^{*}\) & 1.71  & 64.75\(^{*}\)\\
DTC\cite{luo2021semi}                       & 84.71 & 2.07  & 61.04 & 79.14 & 3.15  & 25.81 & 86.26 & 1.99  & 66.39 & 87.50 & 1.54  & 70.10 & 85.31\(^{*}\) & 1.92\(^{*}\)  & 60.69\(^{*}\) \\
ICT\cite{VERMA202290}                       & 86.57 & 1.79  & 68.01 & 80.17 & 2.83  & 40.29 & 85.79 & 2.03  & 58.71 & 88.17 & 1.33  & 72.42 & 86.23\(^{*}\) & 1.77  & 64.81\(^{*}\) \\
MCF\cite{10204429}              & 85.64 & 2.09  & 64.77 & 78.53 & 3.29  & 16.00 & 86.33 & 1.97  & 65.92 & 88.20 & 1.52  & 72.35 & 85.54\(^{*}\) & 2.06\(^{*}\)  & 60.01\(^{*}\) \\
CC-Net\cite{huang2023complementary}         & 85.23 & 1.88  & 63.38 & 78.63 & 2.89  & 33.67 & 86.14 & 1.75  & 66.53 & 87.17 & 1.39  & 69.61 & 85.16\(^{*}\) & 1.79  & 64.81\(^{*}\) \\
\hline
VM\cite{balakrishnan2019tmi}                & - & -  & - & 73.99 & 4.03  & 6.65 & 83.89 & 2.37  & 57.24 & 87.79 & 1.69  & 70.35 & 83.01\(^{*}\)      & 2.46\(^{*}\)      & 42.33\(^{*}\) \\
\hline
T-VOS \cite{yan2023two} & - & -  & - & 80.29 & 2.81  & \textbf{44.54} & 83.97 & 2.66  & 60.63 & 85.15 & 2.05  & 63.98 & 84.56\(^{*}\)      & 2.12\(^{*}\)      & 61.36\(^{*}\) \\
\hline
\textbf{\ourmodel~}              & \textbf{87.42} & 1.83  & \textbf{70.55} & \textbf{82.69} & \textbf{2.45}  & 34.96 & \textbf{87.14} & 2.03  & \textbf{69.14} & \textbf{89.20} & \textbf{1.33}  & \textbf{75.10} & \textbf{87.30} & 1.75  & \textbf{66.71} \\
\Xhline{1.2pt}
\end{tabular}
} 
\end{table}

\begin{table}[t]
\caption{Ablation study of \ourmodel. M and T represent MCL and TCR, respectively. Four key phases denote the four key phases (ES, MD, MD-1, ED). * denotes statistically significant differences (t-test) between the best model and other configurations.}
\label{tab:ablation}
\scalebox{0.66}{
\begin{tabular}{l|ccccccccc|ccccccccc}
\Xhline{1.2pt}
\multirow{3}{*}{Method} & \multicolumn{9}{c|}{Four Key Phases} & \multicolumn{9}{c}{All Phases} \\ \cline{2-19} 
                        & \multicolumn{3}{c|}{AL} & \multicolumn{3}{c|}{PL} & \multicolumn{3}{c|}{Mean} & \multicolumn{3}{c|}{AL} & \multicolumn{3}{c|}{PL} & \multicolumn{3}{c}{Mean} \\ \cline{2-19} 
                        
                        & \multicolumn{1}{c|}{\makecell{Dice}}  & \multicolumn{1}{c|}{\makecell{HD}} & \multicolumn{1}{c|}{\makecell{Conf}} & \multicolumn{1}{c|}{\makecell{Dice}}  & \multicolumn{1}{c|}{\makecell{HD}} & \multicolumn{1}{c|}{\makecell{Conf}} & \multicolumn{1}{c|}{\makecell{Dice}}  & \multicolumn{1}{c|}{\makecell{HD}} & \multicolumn{1}{c|}{\makecell{Conf}} & \multicolumn{1}{c|}{\makecell{Dice}}  & \multicolumn{1}{c|}{\makecell{HD}} & \multicolumn{1}{c|}{\makecell{Conf}} & \multicolumn{1}{c|}{\makecell{Dice}}  & \multicolumn{1}{c|}{\makecell{HD}} & \multicolumn{1}{c|}{\makecell{Conf}} & \multicolumn{1}{c|}{\makecell{Dice}}  & {\makecell{HD}} & \multicolumn{1}{|c}{\makecell{Conf}} \\ 
                        \Xhline{1.2pt}
Based            & \multicolumn{1}{c|}{85.81}    & \multicolumn{1}{c|}{1.93}   & \multicolumn{1}{c|}{\textbf{62.16}} & \multicolumn{1}{c|}{83.21}    & \multicolumn{1}{c|}{2.29}   & \multicolumn{1}{c|}{55.55} & \multicolumn{1}{c|}{84.51}    & \multicolumn{1}{c|}{2.11}     & \multicolumn{1}{c|}{58.85} & \multicolumn{1}{c|}{86.95}    & \multicolumn{1}{c|}{\textbf{1.74}}   & \multicolumn{1}{c|}{\textbf{66.98}} & \multicolumn{1}{c|}{84.08}    & \multicolumn{1}{c|}{1.98}   & \multicolumn{1}{c|}{59.53} & \multicolumn{1}{c|}{85.51\(^{*}\)}    & \multicolumn{1}{c|}{1.86}   & \multicolumn{1}{c}{63.25\(^{*}\)}   \\

Based+M                & \multicolumn{1}{c|}{86.90}    & \multicolumn{1}{c|}{1.92}   & \multicolumn{1}{c|}{61.63} & \multicolumn{1}{c|}{85.15}    & \multicolumn{1}{c|}{2.23}   & \multicolumn{1}{c|}{60.42} & \multicolumn{1}{c|}{86.02}    & \multicolumn{1}{c|}{2.08}     & \multicolumn{1}{c|}{61.03} & \multicolumn{1}{c|}{87.75}    & \multicolumn{1}{c|}{1.78}   & \multicolumn{1}{c|}{66.71} & \multicolumn{1}{c|}{85.57}    & \multicolumn{1}{c|}{2.00}   & \multicolumn{1}{c|}{62.12} & \multicolumn{1}{c|}{86.66\(^{*}\)}    & \multicolumn{1}{c|}{1.89}   & \multicolumn{1}{c}{64.41\(^{*}\)}   \\ 

Based+M+T              & \multicolumn{1}{c|}{\textbf{87.64}}    & \multicolumn{1}{c|}{\textbf{1.90}}   & \multicolumn{1}{c|}{60.35} & \multicolumn{1}{c|}{\textbf{85.58}}    & \multicolumn{1}{c|}{\textbf{1.92}}   & \multicolumn{1}{c|}{\textbf{64.52}} & \multicolumn{1}{c|}{\textbf{86.61}}    & \multicolumn{1}{c|}{\textbf{1.91}}    & \multicolumn{1}{c|}{\textbf{62.43}}  & \multicolumn{1}{c|}{\textbf{88.41}}    & \multicolumn{1}{c|}{1.75}   & \multicolumn{1}{c|}{66.82} & \multicolumn{1}{c|}{\textbf{86.19}}    & \multicolumn{1}{c|}{\textbf{1.76}}   & \multicolumn{1}{c|}{\textbf{66.61}} & \multicolumn{1}{c|}{\textbf{87.30}}    & \multicolumn{1}{c|}{\textbf{1.75}}  & \multicolumn{1}{c}{\textbf{66.71}}  \\ \Xhline{1.2pt}
\end{tabular}
}
\end{table}

\textbf{Ablation Study.}
To demonstrate the impact of different components, we performed an ablation study in Table~\ref{tab:ablation}. Based-\ourmodel~means directly setting labeled and unlabeled data from the same patient based on Mean Teacher learning strategy. 
When integrating MCL, the mean Dice improves by \(1.51\%\) (\(84.51\% \rightarrow 86.02\%\)) for all four key phases segmentation. 
Specifically, for AL and PL Dice, notably increasing \(+1.09\%\) and \(+1.94\%\), respectively. This validates MCL's ability to propagate motion cues across phases. TCR further refines anatomical plausibility, reducing PL HD by \(0.31\,\text{mm}\) (\(2.23 \rightarrow 1.92\)) and achieving a mean HD of \(1.91\,\text{mm}\). Under full-phase evaluation, applying TCR improves PL Conf by \(2.3\%\) (\(64.41\% \rightarrow 66.71\%\)), which significantly improves the consistency among different phases. The best model, Based+MCL+TCR, achieves a \(1.79\%\) improvement in Dice, a \(0.11\,\text{mm}\) reduction in HD and a \(3.46\%\) improvement in Conf compared to the baseline. The progressive improvements highlight their complementary roles: MCL captures temporal dependencies through BMB, while TCR enforces surface-volume continuity via physical regularization.

\begin{figure}[!t]
	\begin{center}
    \includegraphics[width=1\columnwidth]{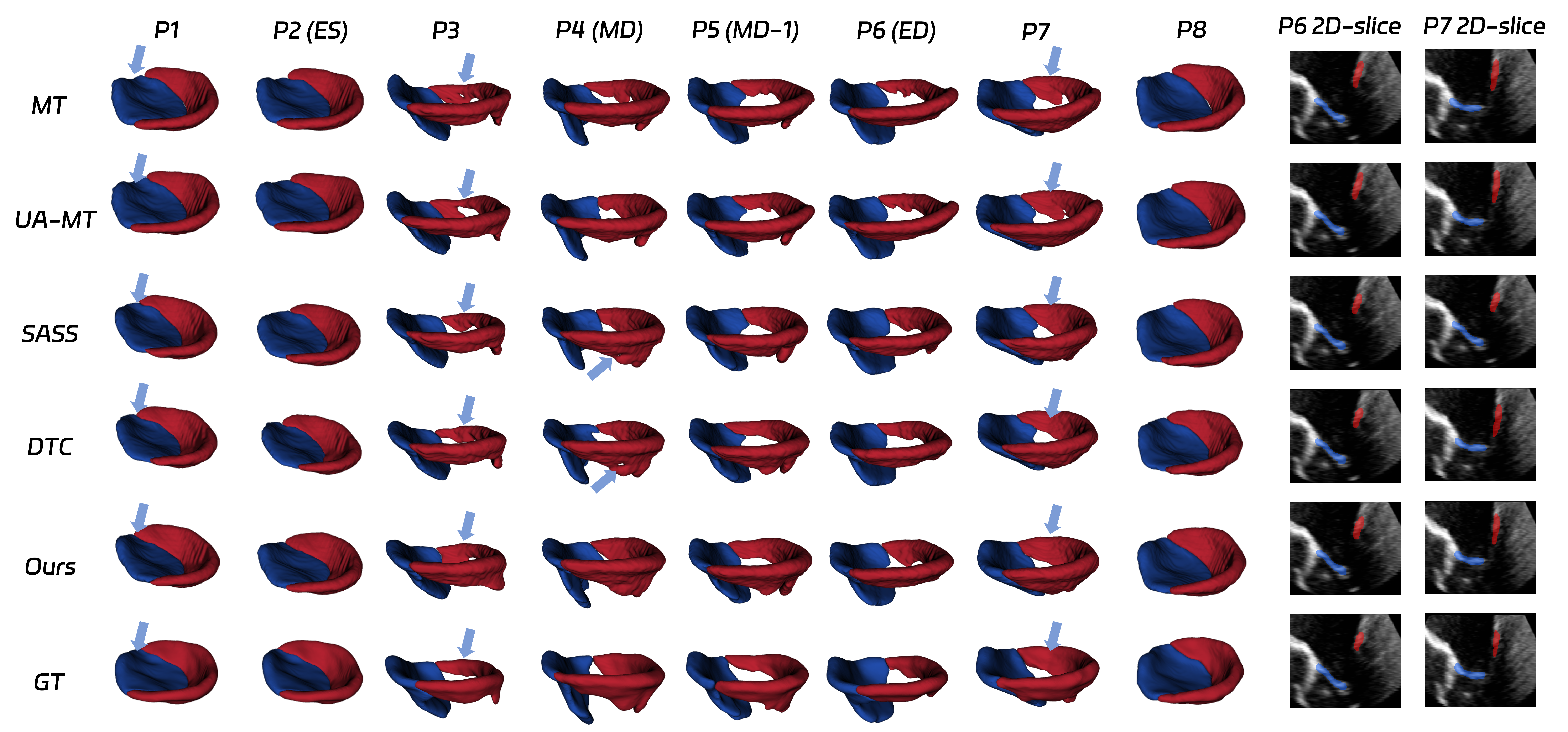}
	\end{center}
    \caption{Examples in consecutive phases with 3D volumes (Left seven columns) and 2D slices (Right two columns). Blue and red represent the AL and PL, respectively.}
	\label{fig:visualization}
\end{figure}

\begin{figure}[!t]
	\begin{center}
    \includegraphics[width=1\columnwidth]{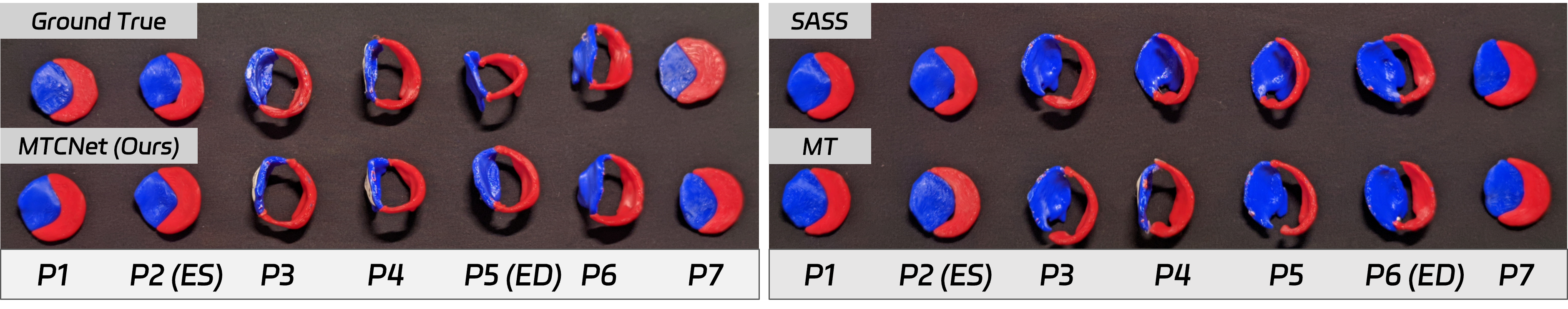}
	\end{center}
    \caption{Examples of 3D printing models of MV among different methods.}
	\label{fig:3Dprint}
\end{figure}

\textbf{Qualitative Results.} 
We further perform a qualitative comparison. 
Fig. \ref{fig:visualization} clearly shows that \ourmodel~produces smoother and more complete segmentations. Particularly in the MD phase, where severe motion and deformation often cause holes (bule arrows).
Artifacts in this region are most prominent due to the valve's motion, indicating \ourmodel~'s superior ability to handle complex motion and structure changes.
To further validate the clinical applicability of our approach, we compared the 3D printing results (See Fig. \ref{fig:3Dprint}). While other methods produce incomplete shapes and missing details, our model achieves greater realism and completeness, showcasing its potential for patient-specific planning.

\section{Conclusion}
In this study, we present \ourmodel, an SSL framework that leverages motion and topology-guided consistency learning for 4D MV segmentation.
\ourmodel~performs better than the SOTA with sparse annotation, especially in motion and structure coherence. We attribute this gain in performance to two key contributions. 
First, the proposed MCL strategy can learn motion information across phases through BMB mechanism, improving segmentation performance in per-phase and inter-phase. 
Additionally, a robust TCR strategy with surface and volume variance as prior knowledge boosts segmentation accuracy while preserving topological coherence. 
\ourmodel~can also facilitate MV dynamics analysis and support personalized hemodynamic assessment and treatment planning.

\begin{credits}
\subsubsection{\ackname}
This work was supported by the grant from National Natural Science Foundation of China (82201851, 82102046, 12326619, 62171290); 
Science and Technology Planning Project of Guangdong Province (2023A0505020002); 
Frontier Technology Development Program of Jiangsu Province (BF2024078);
Guangxi Province Science Program (2024AB17023);
Yunnan Major Science and Technology Special Project Program (202402AA310052);
Yunnan Key Research and Development Program(202503AP140014).

\subsubsection{\discintname}
The authors have no competing interests to declare that are relevant to the content of this article.
\end{credits}

\bibliographystyle{splncs04}
\bibliography{Paper-3656}

\end{document}